\documentstyle[ApJ,times,psfig]{article}

\begin{document}

\def\simg{\mathrel{%
      \rlap{\raise 0.511ex \hbox{$>$}}{\lower 0.511ex \hbox{$\sim$}}}}
\def\siml{\mathrel{%
      \rlap{\raise 0.511ex \hbox{$<$}}{\lower 0.511ex \hbox{$\sim$}}}}
\def\Mesz{M\'esz\'aros~} \def\hsp{\hspace*{2mm}}
\def\ie{i.e$.$~} \def\eg{e.g$.$~} \def\etal{et al$.$~} \def\eq{eq$.$~} 
\def\Lunit{\rm{erg\,s^{-1}}~} \def\Cunit{\gamma\,{\rm cm^{-2} s^{-1}}~}

\title{Analysis of Temporal Features of Gamma-Ray Bursts in the Internal Shock Model}

\author{M. Spada\altaffilmark{1,2}, A. Panaitescu\altaffilmark{1} 
        \& P. \Mesz\altaffilmark{1}}
\affil{Department of Astronomy \& Astrophysics, Pennsylvania State University, 
       University Park, PA 16802}
\altaffiltext{1}{also Institute of Theoretical Physics, 
                 University of California, Santa Barbara}
\altaffiltext{2}{also  Dipartimento di Astronomia e Scienza dello Spazio, 
Universit\`a degli studi di Firenze, Italy }

\begin{abstract}
In a recent paper we have calculated the power density spectrum of Gamma-Ray Bursts 
arising from multiple shocks in a relativistic wind. The wind optical thickness is 
one of the factors to which the power spectrum is most sensitive, therefore we have 
further developed our model by taking into account the photon down-scattering on 
the cold electrons in the wind. For an almost optically thick wind we identify a 
combination of ejection features and wind parameters that yield bursts with an 
average power spectrum in agreement with the observations, and with an efficiency 
of converting the wind kinetic energy in 50--300 keV emission of order 1\%. For 
the same set of model features the interval time between peaks and pulse fluences 
have distributions consistent with the log-normal distribution observed in real 
bursts.
\end{abstract}

\keywords{gamma-rays: bursts - methods: numerical - radiation mechanisms: non-thermal}

\section{Introduction}

The Gamma-Ray Burst (GRB) light-curves are complex and irregular, without any systematic 
temporal features (Fishman \& Meegan 1995) and an understanding of the origin of the 
temporal behavior of GRBs remains an open issue. Statistical studies are
necessary in order to identify the physical properties of the emission mechanism
existent in all or a group of GRBs. Recently Beloborodov et al. 1998, hereafter BSS98, have used
the Fourier analysis of a sample of long GRB light-curves to study the statistical 
properties of their power density spectra (PDS). The PDS features together with other
temporal properties of the observed GRBs, such as the distributions of the time 
interval between peaks and of the pulse fluence (McBreen et al. 1994, 
Li \& Fenimore 1996), can be used to constrain the physical characteristics of 
the GRB source.

In the framework of the internal shock model, the rapid variability and complexity
of the GRB light-curves is due to the emission from multiple shocks 
in a relativistic wind (Rees \& \Mesz 1994, Kobayashi et al. 1997, 
Daigne \& Mochkovitch 1998). The ejecta are released by the source during a time  
comparable to the observed burst duration. The instability of the wind
leads to shocks which convert a fraction of the bulk kinetic energy in internal 
energy at a distance $R \sim 10^{12}-10^{14}$ cm from the central engine.
A turbulent magnetic field is generated and electrons are shock-accelerated, 
leading to synchrotron emission and inverse Compton scatterings.
Within the framework of the internal shock model an alternative hypothesis
about the particle acceleration and radiation emission is the quasi-thermal
Comptonization proposed by Ghisellini \& Celotti (1999), in which particles are 
re-accelerated for all the duration of the collision.
 
In this paper we analyze the features of the GRB light-curves arising
from internal shock model, in order to identify the parameters that
affect most strongly the GRB emission (\S 2). By comparing the features of the
simulated bursts with the observed burst PDS and the distributions of the interval 
time between peaks and of the pulse fluence, we constrain some of the physical 
properties of the ejecta.

\section{Outline of the Model}

We simulate GRB light-curves by adding pulses radiated in a series of 
internal shocks that occur in a transient, unstable relativistic wind. 
As we showed in PSM99 the observed burst variability time-scale depends mostly on
the wind dynamics, its optical thickness and its radiative efficiency
in the BATSE window. Here we model the wind dynamics and the emission processes 
as in PSM99, but we include a more accurate treatment of the photon down-scattering 
on the cold electrons in the wind. We calculate the effect of the photon
diffusion through the colliding shells and the wind on the pulse duration 
and on the energy of the emergent photon, rather than just attenuating 
the pulse fluence according to the optical thickness of the wind 
through which it propagates.
However the contribution of these photons to the duration of 
the received pulses may be important for bursts that are not very optically thin, 
and the photon down-scattering should be taken into account for more reliable
calculations of GRB light-curves.  

As described in PSM99, the wind is discretized as a sequence of $N=t_w/t_v $
shells, where $t_w$ is the duration time of the wind ejection from the central 
source and $t_{v}<<t_w$ is the average interval between consecutive ejections. 
The shell Lorentz factors $\Gamma_i$ are random between $\Gamma_m$ and $\Gamma_M$,
where $\Gamma_M$ can be constant during $t_w$ ("uniform wind") or modulated on 
time scale $\siml t_w$ ("modulated wind").  The shell mass $M_i$ is drawn from a 
log-normal distribution with an average value $\overline{M}=M_w/N$ and a dispersion 
$\sigma_M=\overline{M}$, where $M_w$ is the total mass in the wind, allowing thus the 
occasional ejection of very massive shells.  The total mass is determined by 
requiring that $\sum_{i=1}^N M_i\Gamma_i c^2= L_w t_w$ where $L_w$ is the wind 
luminosity. The time interval $\Delta t_i$ between two  consecutive ejections $i$ 
and $i+1$ is proportional to the $i$-th shell energy, resulting in a wind 
luminosity constant throughout the entire wind, and equal to a pre-set value $L_w$.
This implies that more energetic shells are followed by longer "quiet" times, during
which the "central engine" replenishes.

Given the wind ejection, we calculate the radii where internal collisions take place 
and determine the emission features for each pulse: observer frame duration, fluence, 
and photon arrival time $T_{ob}$, accounting for relativistic and cosmological
effects. The peak photon flux for each pulse is calculated assuming the
pulse shape that Norris \etal (1996) identified in the real bursts, described by
a two-sided exponential function. The addition of all the pulses, as seen by the 
observer in the 50--300 keV range (the $2^{\rm nd}$ and $3^{\rm rd}$ BATSE channels) 
gives the burst $\gamma$-ray light-curve, that is binned on time-scale of 64 ms and 
is used for the computation of the power spectrum. 

For each collision there is a reverse (RS) and a forward shock (FS). The shock 
jump equations allow the calculation of the physical parameters of the shocked 
fluids (Panaitescu \& \Mesz 1999), determine the velocity of these shocks $v_{sh}$,
the compression ratio, the thickness $\Delta$ of the merged shell at the end of 
the collision, and the internal energy in the shocked fluid $U'_s$ (primed quantities 
are measured in the co-moving frame). The accelerated electrons
- a fraction $\zeta_e$ of the total number - have a power-law distribution
of index $-p$, starting from a low Lorentz factor $\gamma_m$. Assuming
that the energy stored in electrons is a fraction $\epsilon_e$ of
the internal energy, we calculate $\gamma_m$ (see PSM99).
The magnetic field $B$ is parameterized through the fraction $\epsilon_B$ of the 
internal energy it contains: $B^2=8\pi\epsilon_B U'_s$.
We assume that between two consecutive collisions the thickness of the shell increases
proportionally to the fractional increase of its radius $d\Delta/\Delta\propto dR/R$.
The shell internal energy increases in each collision by the fraction of $U'_s$ 
that is not radiated, and decreases during the expansion due to adiabatic losses.

The shock-accelerated electrons radiate and the emitted photons can be up-scattered
on the hot electrons ($\gamma_e >> 1$) or down-scattered by the cold ones ($\gamma_e \approx 1$). 
Far from the Klein-Nishina regime the optical depth to up-scattering is $\tau_{ic}=\sigma_{Th} 
\zeta_e n_e' \min (ct'_{\gamma},\Delta')$, where $n'_e$ is the co-moving electrons density 
and $t'_{\gamma} = t'_{sy}/(1+y)$ is the radiative time scale, with $t'_{sy}$ 
the synchrotron cooling time and $y$ the Comptonization parameter (for $\tau_{ic} < 1$,
$y=\gamma_m^2\tau_{ic}$). The optical thickness $\tau_c$ for the cold electrons within
the emitting shell is evaluated by taking into account the cold electrons within the hot fluid, 
those that were accelerated but have cooled radiatively while the shock crossed the shell, 
and those within the yet un-shocked part of the shell. 

A fraction $\min (1,\tau_{ic})$ of the synchrotron photons is inverse Compton scattered 
$n_{ic}= \max (1,\tau_{ic}^2)$ times, unless the Klein-Nishina regime is reached.
The energy of the up-scattered photon and the ratio of the Compton to synchrotron power 
can be cast in the forms:
\begin{equation}
 h\nu_{ic} = \min \left[ \gamma_m m_e c^2,\left( \frac{4}{3} \gamma_m^2
             \right)^{n_{ic}} h\nu_{sy}\right]  \;,
\label{nuic}
\end{equation}
\begin{equation}
 \frac{P_{ic}}{P_{sy}} = \min \left\{ \gamma_m \frac{m_ec^2}{h\nu_{sy}},
    \left[\frac{4}{3}\gamma_m^2 \min (1,\tau_{\zeta}) \right]^{n_{ic}} \right\} \;,
\end{equation}
which take into account the upper limits imposed by the Klein-Nishina effect.
Figure 1$c$ shows the evolution of the synchrotron and inverse Compton peak energies 
during the wind expansion: the energy is lower for larger collisions radii, due to 
the increased shell volume and the less relativistic shocks, which lead to 
lower magnetic fields and electron random Lorentz factors.

The duration $\delta T_0$ of the emitted pulse (\ie ignoring the diffusion through 
optically thick shells) is determined by (1) the spread in the photons 
arrival time $\delta T_{\theta} \approx R/(2 \Gamma^2 c)$ due to the geometrical curvature 
of the emitting shell, (2) the shock shell-crossing time $\delta T_{\Delta}=\Delta/|v_{sh}-v_0|$, 
(where $v_0$ is the shell pre-shock flow velocity), and (3) the radiative cooling time 
$\delta T_{\gamma}\approx t_{\gamma}'/\Gamma$, which we add in quadrature to determine 
$\delta T_0$. As shown in Figure 1b, all these time 
scales increase on average with radius: $\delta T_{\theta}$ is proportional to $R$, 
$\delta T_{\Delta}$ increases due to  the continuous widening of the shell, and 
$\delta T_{\gamma}$ is longer for later collisions because $\gamma_m$ and $B$ are lower.
For $\zeta_e = 1$, $\epsilon_e \approx 0.25$ and $\epsilon_B \approx 0.1$ the radiative 
cooling time is negligible respect to $\delta T_{\theta}$ and $\delta T_{\Delta}$ for 
collisions occurring  at $R < 5 \times 10^{14}$ cm, while for larger radii 
$\delta T_{\gamma}$ is the dominant contribution to the pulse duration (Figure 1b).
For the assumed linear shell broadening between consecutive collisions we find numerically 
that the angular spread and shock-crossing times are comparable during the entire wind 
expansion.

The optical thickness $\tau_c$ is mainly determined by the wind luminosity $L_w$
and the range of Lorentz factors in the wind. In Figure 1$a$ for $30 < \Gamma < 1000$ and 
$L_w=10^{53}\,\Lunit$ most collisions occur at $R = 5 \times 10^{13}$ 
-- $10^{15}$ cm where the emitting shells are optically thin.
For lower Lorentz factors ($5 < \Gamma < 300$) the collisions take place at smaller 
radii ($R = 10^{12}-10^{13}$ cm) and the wind is optically thick (Figure 1$d$). 
When $\tau_c > 1$ photons are down-scattered by the cold electrons before they escape 
the emitting shell, leading to a decrease in the photon energy and an increase of the pulse 
duration. For down-scatterings occurring in the Thomson limit ($\varepsilon' \ll m_e c^2/\gamma_e$)
the energy of the emergent photon can be approximated by $\varepsilon'_{ds} = \varepsilon'
(1- \varepsilon'/m_e c^2)^{\tau_c^2}$, where $\tau_c^2$ is the average number
of scatterings suffered by a photon. For more energetic photons, 
we evaluate $\varepsilon'_{ds}$ numerically, because the cross section depends on the 
photon energy and changes after each photon-electron interaction.
For the set of parameters considered in this paper, the Thompson limit is usually
a good approximation to treat the down-scattering of the synchrotron photons during
all the wind expansion. For the smaller collision radii the inverse Compton 
emission peaks at large comoving frame energies and the general case has to be considered. 
Figure 1$f$ shows the evolution of the synchrotron and inverse Compton observer frame peak 
energies for a thick wind. At $R \approx 10^{12}$ cm, $\tau_c \approx 10^3$ 
and the $\sim 10$ keV synchrotron emission is  down-scattered by an order of magnitude, 
while $\sim 100$ MeV inverse Compton radiation is down-scattered to $\approx 10$ keV.

We approximate the increase in the pulse duration due to the diffusion through optically 
thick shells by the time $\delta T_d$ it takes to a photon to diffuse through them, 
which we add to $\delta T_0$ to determine the observed pulse duration $\delta T$. 
In the Thompson limit $\delta T_d \approx 5 \tau_c \Delta/(24 c)$; in the general 
case the diffusion time is given by
$\delta T_d = \left\{ \sum_{i=1}^{n_s} [\tau(\varepsilon'_i)]^{-1} \right\} \Delta/(2c)$,
where $\tau(x_i)$ is the optical thickness for the $i$-th scattering
and $n_s$ is the number of down-scatterings on the cold electrons, evaluated 
requiring the photon random walk equal to the shell width.
Figure 1$e$ shows the evolution of the pulse duration during the wind expansion:
for smaller collision radii $\delta T_d > \delta T_0$ and the pulse duration is determined
by $\delta T_d$ which decreases with $R$. For larger radii $\delta T_d < \delta T_0$,
thus $\delta T \approx \delta T_0$ and increases with $R$.

For a given pulse, we add to the pulse duration the diffusion time it takes the photon
to propagate through all the shells of optical thickness above unity.
As shown in Figure 1$d$, the wind optical thickness is 1--2 orders of magnitude smaller than the 
optical thickness of the emitting shell ($\tau_c$). Nevertheless the photon diffusion 
through the optically thick shells in the wind can contribute up to 30\% to the pulse 
duration because of the broadening of the shell width during the wind expansion.

The 30--500 keV pulse energy is a fraction of the kinetic energy of the colliding shells, 
equal to the product of the dynamical ($\epsilon_d$), the radiative ($\epsilon_r$), and the 
window efficiency ($\epsilon_w$). \\
\hspace*{2mm}
1) The {\em dynamical efficiency} is the fraction of the kinetic energy that is converted 
   to internal, and is given by the energy and momentum conservation in the collision of a 
   forward shell ($M_f$, $\Gamma_f$) caught up by a back shell ($M_b$, $\Gamma_b > \Gamma_f$):
  \begin{equation}
   \epsilon_d = 1 - \frac{M\Gamma}{\Gamma_b M_b+\Gamma_f M_f}
  \end{equation}
   where $M=M_b+M_f$ is the total mass and 
  \begin{equation}
   \Gamma = \left[\frac{\Gamma_b M_b+\Gamma_f M_f}{M_b/\Gamma_b+M_f/\Gamma_f}\right]^{1/2}
  \end{equation}
   is the final Lorentz factor of the merged shell.
   The $\epsilon_d$ decreases with  $\Gamma_b/\Gamma_f$ and is maximized by $M_b = M_f$, so 
   the inner collisions, for which the difference in the shells Lorentz factor is larger, 
   are the most dynamically efficient, with $\epsilon_d \simg 0.1$ . During the wind 
   expansion the collisions diminish the initial difference in the Lorentz factors and
   the dynamical efficiency decreases to 1\% or less. As show in the next section,
   a modulation in the ejection Lorentz factor is necessary to dynamically efficient 
   collisions at larger radii. \\
\hspace*{2mm} 
2) The {\em radiative efficiency} is the fraction of the internal energy converted in
   radiation, and is given by:
  \begin{equation}
   \epsilon_r = \epsilon_e \frac{t_{\gamma}^{-1}}{t_{\gamma}^{-1}+t_{ad}^{-1}}
  \end{equation}
   where $t_{ad} \sim R/c$ is the adiabatic time-scale. The radiative efficiency decreases 
   during the wind expansion and it's upper limit is the fraction $\epsilon_e$ of 
   internal energy stored in electrons. For magnetic fields not too far from equipartition,
   the radiative timescale is determined by the synchrotron losses. \\
\hspace*{2mm}
3) The {\em window efficiency} is the fraction of the radiated energy that arrives at 
   observer in the 50--300 keV band. The calculation of $\epsilon_w$ is based on the 
   approximation of the synchrotron spectrum by three power-laws, with breaks at the
   cooling frequency $\nu_c$ and the peak frequency $\nu_{sy}$ (at which the $\gamma_m$-electrons 
   radiate). If the $\nu _c < \nu_{sy}$ then the shape of the spectrum is given by:
  \begin{equation}
    F_{\nu} \propto \left\{ \begin{array}{ll} \nu^{1/3} & \nu<\nu_c \\
       \nu^{-1/2} & \nu_c<\nu<\nu_{sy} \\ \nu^{-p/2} & \nu_{sy}<\nu \end{array} \right. \;,
   \label{Fnu1}
  \end{equation}
   where $p$ is the index of the assumed power-law electron distribution. 
   If $\nu_{sy} < \nu_c$ then
  \begin{equation}
    F_{\nu} \propto \left\{ \begin{array}{ll} \nu^{1/3} & \nu<\nu_{sy} \\
       \nu^{-(p-1)/2} & \nu_{sy}<\nu<\nu_c \\ \nu^{-p/2} & \nu_c<\nu \end{array} \right.\;.
   \label{Fnu2}
  \end{equation}
   The inverse Compton spectrum has the same shape but is shifted to higher energy by
   the factor implied by equation (\ref{nuic}). For optically thick emitting shells we 
   approximate the burst spectrum as given in equations (\ref{Fnu1}) and (\ref{Fnu2}), 
   using the down-scattered cooling and peak frequencies.

\section{Effect of the Physical Parameters on the GRB Power Density Spectrum}

In this section we analyze the effect of the model parameters on two distributions 
that characterize the GRB temporal structure: the PDS and the distribution of the 
interval $\delta_p$ between peaks.  The relevant model parameters describe the wind 
ejection ($t_v$, $t_w$, $\Gamma_{min}$, $\Gamma_{max}$ and $L_w$) 
and the energy release ($\epsilon_e$, $\zeta_e$ and $\epsilon_B$).
In order to diminish the large PDS fluctuations, in this section we use power 
spectra that are averaged over 10 peak-normalized bursts. The light-curve peaks are
identified with the peak finding algorithm (PFA) described by Li \& Fenimore (1996). 
For each time bin $T_p$ with a photon flux $C_p$ higher than those of the neighboring time
bins, we search for the times $T_1 < T_p $ and $T_2 > T_p$ when the photon fluxes $C_1$
and $C_2$ satisfy $C_p - C_{1,2} > N_{var} \sqrt{C_p}$. A peak is identified at $T_p$
there are no time bins between $T_1$ and $T_2$ with photon fluxes higher than $C_p$.
The light-curve valleys are identified as the minima between two consecutive peaks.

The energy release parameters $\epsilon_e$, $\zeta_e$, and $\epsilon_B$ determine 
the 50--300 keV radiative efficiency of the pulses.
In an optically thin wind, the parameter that affects mostly the
window efficiency is the electron injection fraction $\zeta_e$
(for an optically thick wind the photons are down-scattered before they escape
the shells and the window efficiency depends also on $\tau_c$).
Figures 2$a$ and 2$b$ show the PDS and the $\delta_p$ distributions for a thin wind 
with $L=10^{52}\,\Lunit$, $\Gamma_m=30$, and $\Gamma_M=800$, 
and for two different $\zeta_e$ (1 and $10^{-3}$).
In both cases synchrotron emission is the dominant radiative process
the inverse Compton contribution to the total emission being 10\% for
$\zeta_e=1$ and less then 0.1\% for $\zeta_e=10^{-3}$.
For $\zeta_e=1$ the synchrotron emission lies mainly below the BATSE window (Figure 2$c$), 
the window efficiency decreases from shorter to longer pulses, the light-curve 
is formed by pulses with a duration of $\delta T\approx 10^{-2}$ s, and with 
an average difference in the photon arrival time $\Delta T \approx 0.2$ s. Because
$\Delta T \gg \delta T$, the distribution of intervals between peaks
is determined by the pulses arrival times and peaks at 0.1 -- 0.2 s.
If $\zeta_e=10^{-3}$ the synchrotron emission is above the BATSE window for 
$\delta T < 0.3 $s and the window efficiency is maximized for 
$\delta T\approx 0.2$ -- $0.4$ s. The light curve is formed by longer pulses,
the lower frequency power in the PDS increases and the interval time between peaks
shifts to longer time-scale.

The 50--300 keV efficiency of the synchrotron and the inverse Compton emissions
is determined by the strength magnetic field $B$.  While for $\epsilon_B > 0.1$ the 
emission is dominated by synchrotron radiation, for values of the magnetic field
well below equipartition ($\epsilon_B <  0.01$) the burst emission is
dominated by the inverse Compton. Because the shape of the PDS does not change,
we conclude that the PDS is not sensitive to $\epsilon_B$
and the relative contribution of synchrotron and inverse Compton in the light-curve.
 
The ejection parameters determine the dynamics of the wind and the evolution of the pulse
dynamical efficiency $\epsilon_d$. The latter reflects the evolution of the differences 
between the Lorentz factors of a pair of colliding shells. 
The first collisions remove the initial random differences,  
and the merged shells have Lorentz factors near the ejection average value 
$\overline{\Gamma}=(\Gamma_m+\Gamma_M)/2$. If the wind is uniform $\overline{\Gamma}$ is the same
for all the shells, resulting in a steady decrease of $\epsilon_d$ during the wind
expansion. If the range of shell ejection Lorentz factors is variable on time scale of the 
order of $t_w$ (a "modulated" wind), $\overline{\Gamma}$
reflects the initial modulation in $\Gamma_M$ 
and large radii collisions that are dynamically efficient are still possible. 

Figure 3$a$ shows the effect on the PDS of square-sine modulations of the upper limit $\Gamma_M$ 
with periods $P=t_w$ and $P=t_w/4$. The $i$-th shell ejection Lorentz factor is given by
\begin{equation}
  \Gamma_i=\Gamma_m+ a_i \sin^2 \left(\frac{\omega i}{N}\right)(\Gamma_M-\Gamma_m),
\label{sqsine}
\end{equation}
where $a_i$ is a random number between 0 and 1, and $\omega = 2 \pi t_w/P$.
The modulation shifts the power from high to low frequencies, and the magnitude
of this shift depends on the modulation period.
If $P = t_w$ the effect of the modulation for interaction radii less than $\approx 10^{14}$ 
cm (corresponding to $\delta T \approx 1$ s) is negligible and the wind evolves as in the 
uniform case: the $\epsilon_d$ decreases from 5\% to 0.2\% when $\delta T$ increases from
0.01 s to 1 s (Figure 3$c$). For $R \simg 10^{14}$ cm the modulation becomes relevant: 
the wind is formed of groups of few massive shells with different Lorentz factors. 
The dynamical efficiency remains constant for subsequent collisions between massive shells,
which yield long pulses ($\delta T = 0.3$ -- $10$ s) that carry a 
substantial fraction of the total burst fluence.

Figure 3$d$ shows that the dependence on $\delta T$ of the synchrotron efficiency $\epsilon_{sy}$ 
of the FS pulses has a similar behavior as that of $\epsilon_d$, because the internal energy 
density in the shocked plasma depends on $\epsilon_d$. For an higher internal energy, 
the minimum electron Lorentz factor $\gamma_m$ increases, leading to a higher energy
emission and a shorter radiative cooling time-scale. Therefore  
the synchrotron efficiency remains constant on the same range of
$\delta T$ where is constant the dynamical efficiency, contributing to a 
shift of power to low frequencies in the PDS.

The optical thickness of the wind depends mostly on the range of shell Lorentz factors 
($\Gamma_m-\Gamma_M$) and on the wind luminosity ($L_w$).  Figure 4$a$ shows PDSs for two 
ranges of Lorentz factors, 30--1000 and 10--150. 
In the former case the wind is essentially optically thin, 
and the photon diffusion does not affect the pulses duration $\delta T$, that  
increases with $R$ (Figure 4$c$) from 0.01 s to 1 s, between
$10^{13}$ and $10^{15}$ cm. In the latter case the wind is optically thick,
in 80\% of the collisions $\tau_c>1$, and the pulse duration is given by the
diffusion time: $\delta T_d$ decreases from $\approx 5$ s
to $\approx 0.6$ s between $R = 3 \times 10^{12}$ cm and $R = 10^{13}$ cm, where 
$\delta T$ is determined mainly by the shell curvature and thus increases with $R$.
For the optically thick wind the long pulses are generated at smaller $R$ where 
the efficiency has the maximum value, and the pulse energy increases with 
$\delta T$ (Figure 4$d$). 
The PDS has more power at low frequency and the time intervals between peaks 
are longer than in the optically thin case (Figure 4$b$).
The 50--300 keV efficiency is of the same order for the two cases:
$4 \times 10^{-3}$ and $5 \times 10^{-3}$ for a range of Lorentz factors of
30--1000 and 10--150, respectively.  

An increase in the wind luminosity has a similar effect on the PDS shape as a decrease in
$\Gamma_m$ and $\Gamma_M$. In the latter case the wind becomes thicker because the shells are 
more massive.

The variability time scale $t_v$ affects the dynamical evolution of the shells in the
following way. If the time intervals between successive ejections delays decreases then the 
collisions occur at smaller radii, where the wind is more optically thick.
The differences between the Lorentz factors diminish faster (there are more shells for 
smaller $t_v$), reducing the dynamical efficiency for short pulses. 
For the modulated wind this effect is more relevant than in the
random case. The duration $t_w$ of the wind ejection determines mainly the number of shells, 
and changes in $t_w$ do not affect much the evolution of uniform winds.
However, for a modulated wind, $t_w$ also determines the number of periods in the
Lorentz factor (if the duration of a period is independent of $t_w$), influencing
thus the clumping of shells.

The burst redshift determines the co-moving energy range which is
redshifted into the observing range, leading to a change in the total pulse efficiency,
and altering the observed pulse duration. Obviously, by increasing the burst redshift, 
power is shifted from higher to lower frequencies.

\section{Comparison with the Observations}

An analysis of the PDS of real bursts was presented by BSS98. They calculated the Fourier 
transform of 214 long ($T_{90} > 20$ s) and bright burst, and have found that
the average PDS is a power-law ($P_f \propto f^{-5/3}$, $f$ is frequency) over almost two 
orders of magnitude in frequency, between 0.02 Hz and 2 Hz, where a break is observed,
indicating a paucity of pulses with duration less than $\approx 0.5$ s. The distribution 
of intervals between peaks has been studied by McBreen et al. (1994) and by 
Li \& Fenimore (1996), who showed that the distributions of the pulse fluence $S_p$ and 
of the time interval $\delta_p$ between peaks are consistent with a log-normal 
distribution.   

As was shown in the previous section, if the wind is optically thin and the
ejection features are random, the pulse duration increases
with the collision radius and the emission efficiency decreases during
the wind expansion. The short inner collisions yield most of the 50--300 keV burst emission
and the internal shock model predicts a flat PDS with equal power at
low and high frequency. Thus, in order to explain the observed behavior,
we need a configuration of the parameters which shifts power from the short to the
long time-scales in the light-curves. Moreover, the $\delta_p$ distribution is not log-normal:
Figures 2$b$, 3$b$, and 4$b$ show that in GRBs arising from optically thin, uniform winds
there are too many short intervals between peaks respect to a Gaussian $\log \delta_p$ 
distribution.  

PSM99 have identified three possible ways to explain the deficit of pulses with $\delta T < 1$ s: \\
\hspace*{2mm}
(1) a reduction in the electron injection fraction. This increases the photon
   energy, reducing the window efficiency of the short pulses (causing the high energy break)
   and increasing that of the longer ones.  However the behavior of the PDS at lower frequency 
   remains flat (see Figure 2$a$). \\
\hspace*{2mm}
(2) a modulation of the shell ejection Lorentz factor. This allows different
   configurations for the collisions series and a higher dynamical efficiencies for
   longer pulses (see Figure 3$a$). \\
\hspace*{2mm}
(3) an increase of the optical thickness of the wind. In this case the down-scattering suffered 
   by the photons as they propagate through the wind increases the pulse duration for the small radii
   collisions, which yield the shorter duration pulses (see Figure 4$a$).

In Figure 5$a$ we show a simulated light-curve for a square-sine modulated wind (with $P=t_w$)
The burst 50--300 keV efficiency is 1\%, and the 90\% of the RS and 80\% of the FS 
propagate in optically thick shells. 
If $N_{var} = 0.1$ (the free parameter of the PFA), we 
find 22 pulses in the light-curve shown in Figure 5$a$. In order to have more peaks we simulate 
four light-curves with the same injection features and wind parameters and we calculate
the interval between peaks $\delta_p$ (Figure 5$b$) and peak fluence $S_p$ (Figure 5$c$)
distributions. The distributions are similar to a log-normal one, and the choice of $N_{var}$
does not affect strongly their shape.

In order to compare the PDS of the simulated bursts with the observed one, we consider 
an ensemble of cosmological GRBs.  Some authors (Totani, 1997, Wijers, \etal 1998, 
Krumholz et al. 1998, Hogg \& Fruchter 1999, Mao \& Mo 1999 ) have used 
a GRB co-moving rate density proportional to the star formation rate. Others 
(Reichart \& \Mesz 1997) have employed a power-law GRB density evolution with redshift, 
which was found by (Bagot et al. 1998) to be consistent for $z \siml 2$ with their 
results from population-synthesis computations of binary neutron stars merger rates. 
Finally, other researchers (Krumholz et al. 1998, Hogg \& Fruchter 1999), have considered 
a constant GRB rate density. In this work, we use the power-law with redshift GRB density 
evolution $n_c(z) \propto (1+z)^D$, mainly as a convenient parameterization. 
An $n_c(z)$ proportional to the star formation rate would lead to different 
sets of model parameters (see below), but the differences are minor, because the two
functions differ substantially in shape only for $z > 1$, where there is a strong decrease 
of the co-moving volume per unit redshift and a smaller chance of obtaining a burst that 
has a 50--300 keV peak photon flux below $1\,\Cunit$ (bursts dimmer than this limit are
not included in the calculation of the average PDS and intensity distribution).

Given the rate density evolution, the GRB redshift is chosen from a probability
distribution
\begin{equation}
 \frac{d{\cal P}}{dz} \propto \frac{n_c(z)}{1+z} \frac{dV}{dz} \;,
\label{dpdz}
\end{equation}
where $dV/dz$ is the cosmological co-moving volume per unit redshift
\begin{equation}
 \frac{dV}{dz} = 4\pi \left(\frac{c}{H_0}\right)^3 \frac{[q_0z-(1-q_0)
                 (\sqrt{2q_0z+1}-1)]^2} {q_0^4 (2q_0z+1)^{1/2} (1+z)^6} \;.
\label{dVdz}
\end{equation}
We assume $q_0=0.5$ and $H_0=75\,{\rm km\, s^{-1} Mpc^{-1}}$.

The inferred isotropic 50--300 keV luminosities of the GRBs that have measured redshifts span 
more than one order of magnitude, therefore the standard candle approximation is not a good 
approximation. We use an un-evolving power-law distribution for the wind luminosity:
\begin{equation}
 \Phi(L) \propto L^{-\beta}\;, \quad L_m \leq L \leq L_M \;,
\label{phiL}
\end{equation}
and zero otherwise. Note that this not the same as assuming that GRBs have a power-law
distribution of their 50--300 keV luminosities, as it is usually done (\eg  Reichart \& 
\Mesz (1997), Krumholz et al. 1998, Mao \& Mo 1999), as the relationship between
the wind and the 50--300 keV luminosities is set by the window efficiency (at the source)
and, in the case of winds that are optically thick, by the wind optical
thickness, both of which are dependent on the wind luminosity.

In finding model parameters that yield bursts consistent with the observations, we held 
constant $t_v=25$ ms, $L_M/L_m=100$, $\beta=2$, $\varepsilon_e=0.25$, $\varepsilon_B=0.1$, 
and $p=2.5$ . The chosen $t_v$ is short enough to ensure that the observed 2 Hz PDS break 
frequency is below $[(1+z) t_v]^{-1}$ (bursts with $z > 3$ are rarely brighter than 
$1\,\Cunit$), corresponding to the pulses that are partly suppressed by the choice of $t_v$. 
the PDS frequencies affected by the assumed $t_v=25$ ms are $\simg 10$ Hz. This may
suggest a possible explanation for the PDS break observed by BSS98: the lack
of pulses shorter than $\sim 1$ s is due to the existence of a minimum wind variability
time-scale of the same order. However, such $t_v$'s would be much larger than the dynamical 
time-scales of plausible GRB progenitors (\Mesz et al 1999), and we do not consider this
a viable possibility. The choices of $L_M/L_m$ and $\beta$ are consistent with the values
found by Reichart \& \Mesz (1997), Mao \& Mo (1999), and Krumholz \etal (1998) from
fits to the observed intensity distribution.  The values chosen for
$\varepsilon_e$ and $\varepsilon_B$ are not too far from those determined by Wijers
\& Galama (1999) from the emission features of the afterglows of GRB 970508 and 971214.
The above value of electron index $p$ is close to the values implied by the observed
slopes of the afterglow optical decays.

Figure 6$b$ shows a burst-averaged PDS whose features are similar to that found by BSS98
in real bursts. The wind ejection is modulated by a square sine (\eq [\ref{sqsine}]) with 
a random period between $t_w/4$ and $t_w$. About 40\% of the 300 simulated bursts
have peak photon fluxes brighter than $1\,\Cunit$.
Taking into account that the average redshift for these bursts is $\overline{z}=0.90$
the average burst duration $\overline{T}_b \approx 1.5 (1+\overline{z}) t_w$ is close
to the value $\overline{T}_b=80$ s of the bursts used by BSS98 (the factor 1.5 was determined
numerically and represents the ratio between the burst duration at the source redshift and
$t_w$). As can be seen in Figure 6$b$, $P_f \propto f^{-5/3}$ between 0.04 Hz and 2 Hz
and falls off steeper at frequencies larger than 2 Hz.
The model parameters that led to the PDS of Figure 6$b$ yield bursts whose integral
intensity distribution is shown in Figure 6$a$, which consistent with the distribution
found by Pendleton \etal (1996): excluding the bursts dimmer than $1\,\Cunit$, the  
model has $\chi^2=9.5$ for 9 degrees of freedom.

\section{Conclusion}

 We have calculated power density spectra of GRBs arising from internal shocks in an 
unsteady relativistic wind. By studying how the features of these spectra depend on
the model parameters (Figures 2, 3, and 4), we have identified a set parameters 
(Figure 6) that leads to bursts whose average PDS exhibits an $f^{-5/3}$ behavior
(where $f$ is frequency) for $0.04\,{\rm Hz} < f < 2\,{\rm Hz}$, as found by BSS98 
in real GRBs. Moreover, the integral intensity distribution of the simulated bursts 
is consistent with that observed by Pendleton \etal (1996), and the distributions of 
the time intervals between peaks and of the pulse fluences are consistent with the 
log-normal distributions identified by Li \& Fenimore (1996) in real bursts.

 The characteristics of the modeled bursts with the above mentioned features
are: (1) a sub-unity electron injection fraction, required to increase the radiative
efficiency of the larger collision radii,
(2) a modulated Lorentz factor of the ejected shells, necessary 
to increase the dynamical wind efficiency during the wind expansion and,  
(3) a shells optical thickness to scattering on cold electrons above unity, required to 
increase the duration of the pulses as they propagate through the colliding shells 
and the wind.

In the internal shock model, the most efficient collisions, with a dynamical 
efficiency of 10-20\% and a radiative efficiency of $10-30\%$, happen in 
the first part of the wind expansion where the wind optically thickness 
is higher and the angular spread time, the shell shock-crossing time and 
the electrons cooling time are shorter ($\ll$ 0.5 s). 
In order to reproduce the observed break at 2 Hz in the PDS, we have previously (see PSM99)
attenuated the fluence of these short pulses according to an high wind optically
thickness, with a resulting low burst efficiency (10$^{-4}$ for an uniform wind
and 10$^{-3}$ for a modulated one). 
The study of the photon diffusion, presented here, allowed us to find model 
parameters that yield an 1\% efficiency of converting the wind kinetic 
energy into 50--300 keV emission. For an optically thick wind, the pulse 
duration of the first, efficient collisions at small radii
is determinated by the time the photons take to escape the shells, that depends only
on the colliding shells width and optically thickness $\tau_c$. If $\tau_c\gg 1$
the diffusion time for the efficient collisions is $\simg$ 0.5 s and the simulated
average PDS shows the break at 2 Hz with a burst efficiency close to the
maximal value (few \%) admitted by the model (see also Kumar 1999).

\acknowledgements{This research is supported by NASA NAG5-2857, NSF PAY94-07194 and the CNR. 
We are grateful to Martin Rees, Stein Sigurdsson and Marco Salvati for stimulating comments.}

\clearpage

\begin{figure*}
\vspace*{0.1cm}
\centerline{\psfig{figure=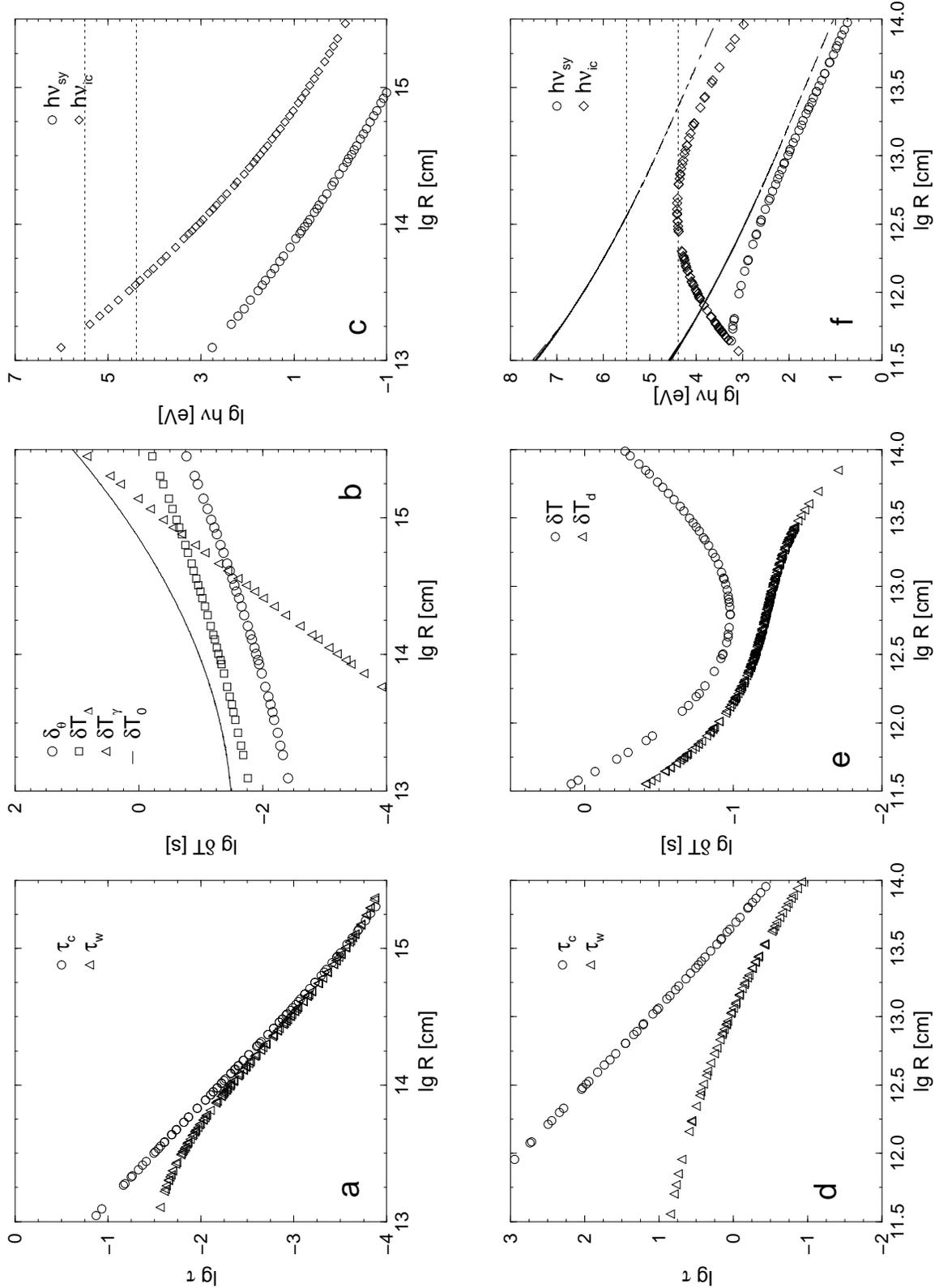}}
\caption{\footnotesize{
Upper panels: optically thin wind with $L_w = 10^{53}\,\Lunit$
and $30 < \Gamma < 1000$. Panel $a$ shows the dependence on the collision
radius $R$ of the optical thickness $\tau_c$ for scattering on the cold electrons inside
the emitting shell, and that of the rest of the wind ($\tau_w$).
Panel $b$ illustrates the $R$-dependence of the pulse duration and of
the terms that contribute to it; while graph $c$ shows the $R$-dependence
of the synchrotron and inverse Compton peak energies.
Lower panels: optically thick wind $L_w=10^{53}\,\Lunit$ and $5 < \Gamma < 300$.
Graph $d$ shows the R-dependence of $\tau_c$ and $\tau_w$,
graph $e$ shows the evolution of the pulse duration $\delta T = \delta T_d +
(\delta T_{\theta}^2 + \delta T_{\Delta}^2+ \delta T_{\gamma}^2)^{1/2}$ and the
contribution of the diffusion time through the wind $\delta T_d$
(\ie excluding the emitting shell).
Graph $f$ shows the down-scattered synchrotron and inverse Compton
energy peaks versus the collision radius.
The dashed lines in panel $f$ show the evolution of the synchrotron
and inverse Compton peak before the down-scattering, and
the dotted lines in panel $c$ and $f$ show the BATSE window.
Parameters: $t_v=0.02$ s, $t_w=20$ s, $\epsilon_e=0.25$, $\epsilon_B=0.1$, $\zeta_e=1$,
burst redshift $z=1$.
Only a small fraction of the total number of pulses is shown; the density of
the points illustrates the radius distribution. The curves shown are log-log
space fits for the most efficient pulses. The actual values are scattered
around the fit. \label{fig1}}}
\end{figure*}

\begin{figure*}
\vspace*{0.1cm}
\centerline{\psfig{figure=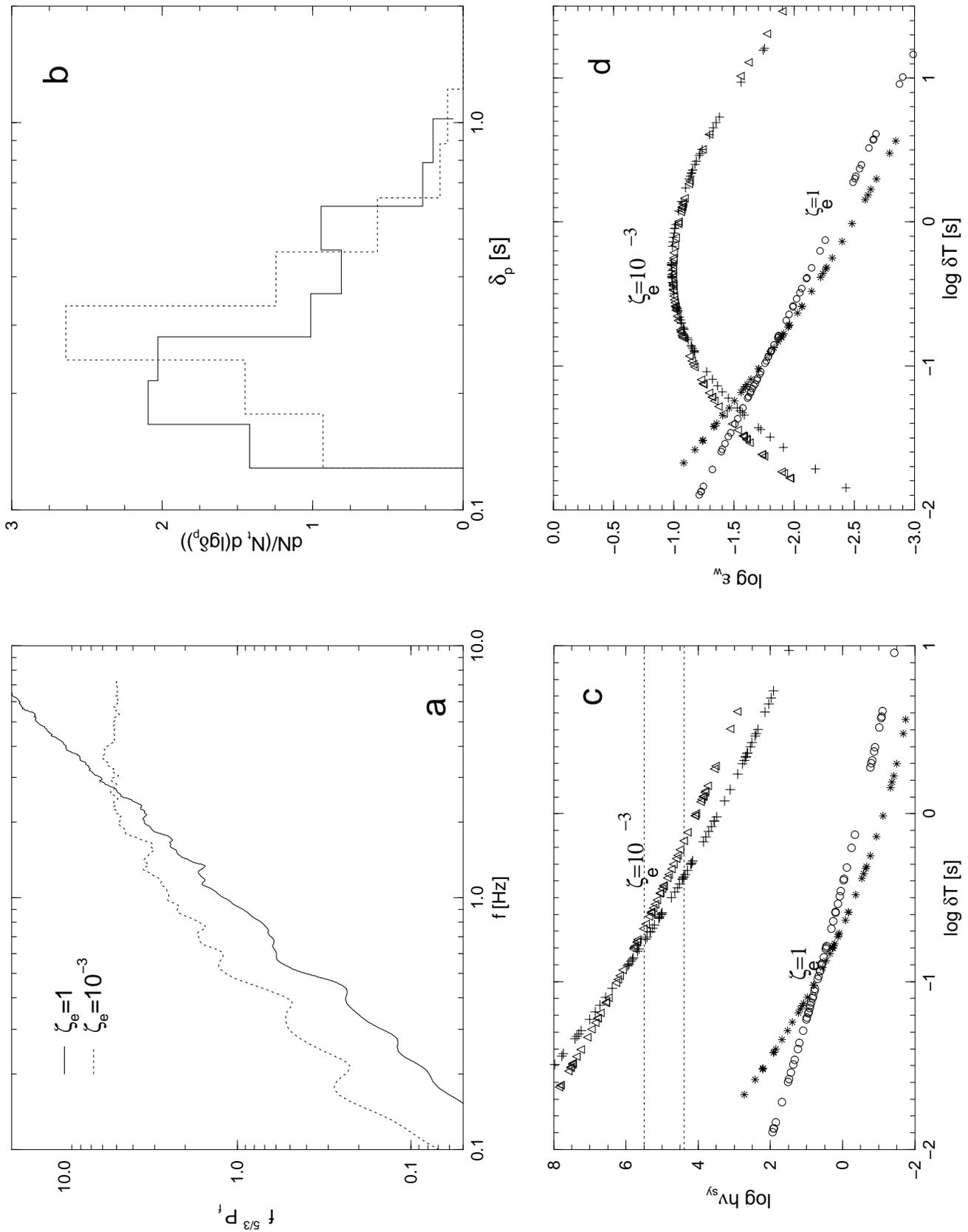}}
\caption{\footnotesize{
     Effect of the injection fraction $\zeta_e$ on the burst PDSs (panel $a$), 
     distribution of interval between peaks (panel $b$) for $N_t \sim 130$ peaks, 
     energy of the synchrotron spectrum peak (panel $c$), and window efficiency (panel $d$).
     Parameters: $L_w=10^{52}\,\Lunit$, $\Gamma_m=30$, $\Gamma_M=800$,
     other parameters are as for Figure 1.
     Panels $c$ and $d$ show log-log space polynomial fits, illustrating thus only the trends. 
     The RS emission is represented with circles and triangles, while the FS one is shown 
     with crosses and stars. \label{fig2}}} 
\end{figure*}

\begin{figure*}
\vspace*{0.5cm}
\centerline{\psfig{figure=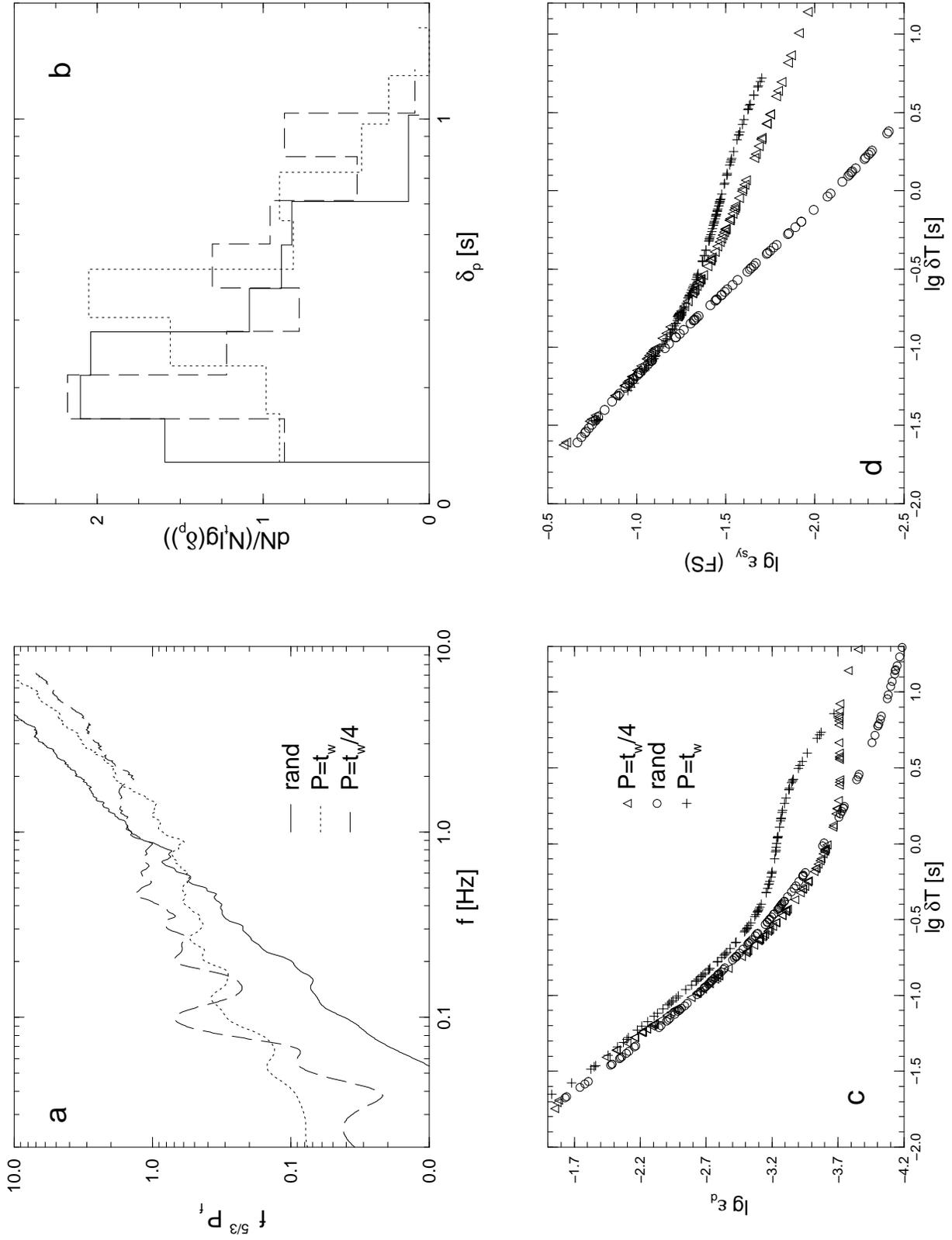}}
\caption{\footnotesize{
    Effect of a modulation of the wind ejection. Panels $a$ and $b$ show a comparison
    between the PDSs and the $\delta_p$ distributions for 
    an uniform wind (solid line), and a square-sine modulated ejection with period 
    $t_w$ (dotted line) and $t_w/4$ (dashed line). Parameters: $L_w=10^{52}\,\Lunit$, 
    $\Gamma_m=30$, $\Gamma_M=1000$, other parameters are as for Figure 2. 
    Lower panels: the dependence of the dynamical efficiency (panel $c$),
    and FS synchrotron efficiency (panel $d$) versus the pulse duration. 
    \label{fig3}}}
\end{figure*}

\begin{figure*}
\vspace*{0.1cm}
\centerline{\psfig{figure=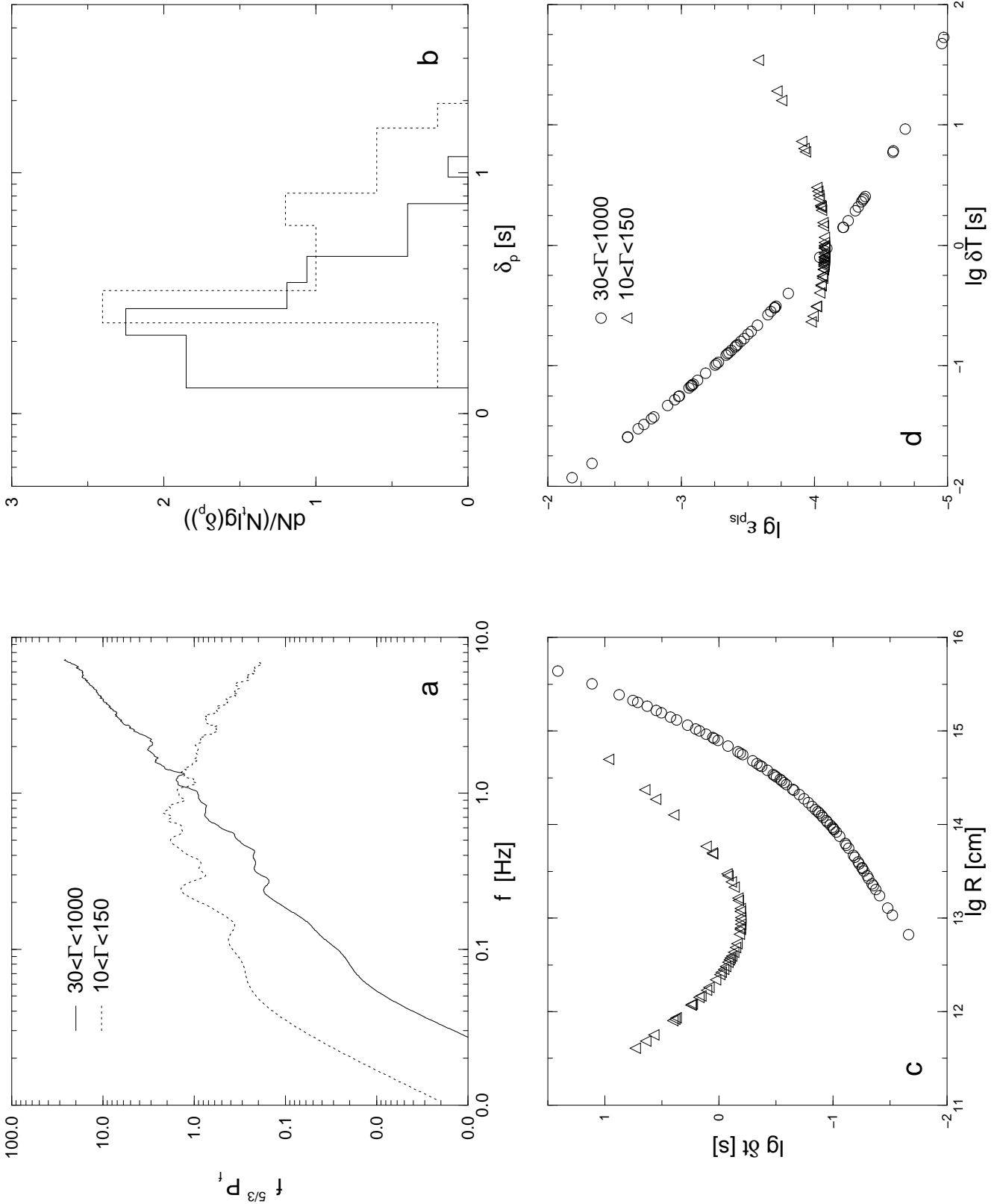}}
\caption{\footnotesize{
    Effect of the optical thickness. Panels $a$ and $b$ show a comparison
    between the PDSs and the distributions of $\delta_p$ for an optically thin 
    wind with $30 < \Gamma < 1000$ ($N_t=70$ peaks), and an optically thick 
    one with $10 < \Gamma < 150$ ($N_t=40$ peaks). 
    $L_w=10^{53}\,\Lunit$, other parameters are as for Figures 2 and 3.
    The dependence of $\delta T$ on the collision radius and of the pulse 50--300 keV 
    fluence on $\delta T$ are shown in panels $c$ and $d$ respectively,
    for the optically thick (triangle) and thin (circle) winds.
    The pulse fluence is the fraction of the total fluence in the light curve carried by
    each pulse. \label{fig4}}}
\end{figure*}

\begin{figure*}
\vspace*{0.1cm}
\centerline{\psfig{figure=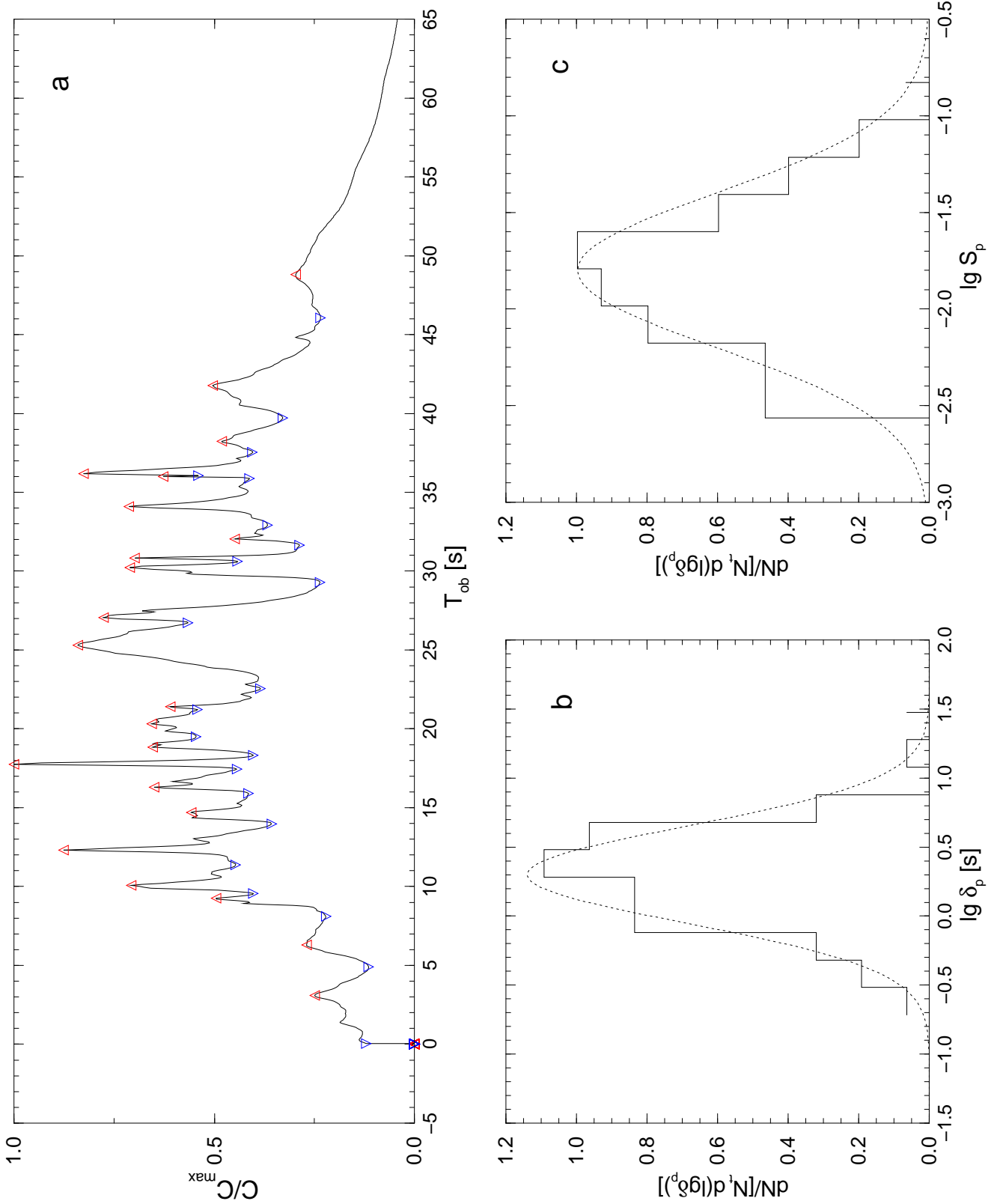}}
\caption{\footnotesize{
    Top panel: light-curve for a modulated wind with $P = t_w$
    and $\Gamma_m=5$, $\Gamma_M=150$, $L_w= 2 \times 10^{52}\,\Lunit$,
    $t_v=25$ ms, $t_w=30$ s, $z=1$, $\zeta_e = 0.1$, $\epsilon_e = 0.25$, $\epsilon_B = 0.1$ .
    The photon flux is normalized to its highest value.
    The triangle the peaks (38) and the valleys selected with the PFA ($N_{var}=0.1$). 
    Bottom: the distributions of the interval $\delta_p$ between peaks ($b$) and of
    the peak fluence $S_p$ ($c$), the latter representing the fraction of the total fluence 
    between two consecutive valleys.
    The distributions are calculated for four bursts with identical ejection features
    and wind parameters, total numbers of peaks is 80. The log-normal distributions
    that fit best the numerical results are shown with dotted curves. \label{fig5}}}
\end{figure*}

\begin{figure*}
\vspace*{0.1cm}
\centerline{\psfig{figure=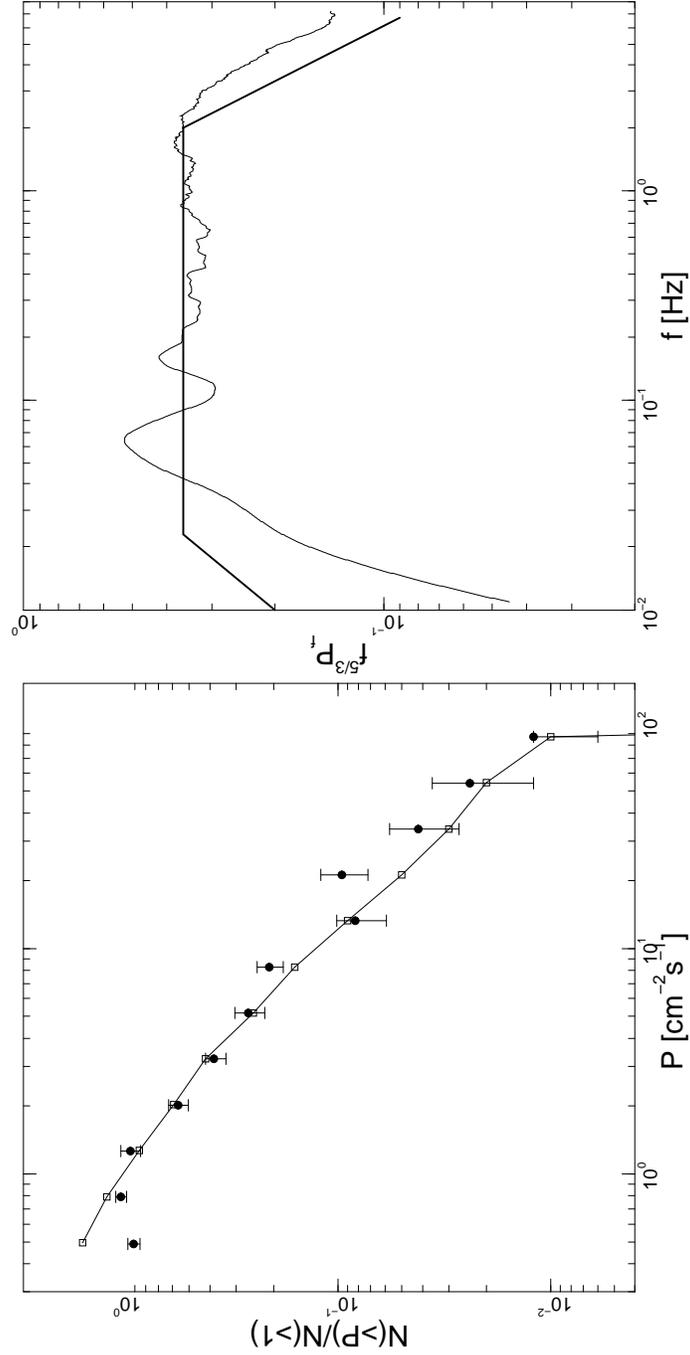}}
\caption{\footnotesize{
    Left panel: intensity distribution of simulated bursts (open squares) compared 
    to that of the observed ones, taken from Pendleton \etal (1996) (filled circles). 
    Both distributions are normalized to the number of bursts brighter than $1\,\Cunit$.
    Right panel: the averaged PDS of the simulated bursts, 
    compared with the shape of the PDS (thick line) determined by BSS98 for real bursts.
    The average is done over 112 bursts with a square-sine modulated wind of random period 
    between $t_w/4$ and $t_w$.
    Parameters: $L_m=2 \times 10^{52}\,\Lunit$, $L_M=100 L_m$, $\beta=2$, $D=3$, 
    $t_w=30$ s, $t_v=25$ ms, $\epsilon_e=0.25$, $\epsilon_B=0.1$, $\zeta_e=0.1$, $\Gamma_m=5$, 
    $\Gamma_M=150$, $D=3$. \label{fig6}}}
\end{figure*}

\end{document}